\documentclass[12pt]{article}

\usepackage{amsmath}
\usepackage{graphicx}
\usepackage{color}
\usepackage[compat=1.0.0]{tikz-feynman}
\usepackage{feynmp-auto}
\usepackage[colorlinks=true, urlcolor=blue, linkcolor=blue, 
citecolor=blue]{hyperref}
\usepackage{cite}

\def \lsim{\mathrel{\vcenter
     {\hbox{$<$}\nointerlineskip\hbox{$\sim$}}}}
\def \gsim{\mathrel{\vcenter
     {\hbox{$>$}\nointerlineskip\hbox{$\sim$}}}}

\newcommand{\vrel}{v_{\rm rel}}
\newcommand{\mred}{\mu_{1A}}
\newcommand{\beq}{\begin{equation}}
\newcommand{\eeq}{\end{equation}}
\newcommand{\beqa}{\begin{eqnarray}}
\newcommand{\eeqa}{\end{eqnarray}}
\newcommand{\beqar}{\begin{eqnarray*}}
\newcommand{\eeqar}{\end{eqnarray*}}

\begin{tiny}

\end{tiny} 

\begin{document}
\thispagestyle{empty}
$\,$

\vspace{32pt}

\begin{center}

\textbf{\Large Monochromatic neutrinos from \\ scotogenic dark matter}

\vspace{50pt}
Ricardo Cepedello, Pablo de la Torre, Manuel Masip
\vspace{16pt}

\textit{Departamento de F{\'\i}sica Te\'orica y del Cosmos}\\
\textit{Universidad de Granada, E-18071 Granada, Spain}\\
\vspace{16pt}

\texttt{ricepe,pdelatorre,masip@ugr.es}

\end{center}

\vspace{30pt}

\date{\today}

\begin{abstract}

The scotogenic model defines a  
framework for radiative neutrino masses and 
provides a viable dark matter candidate. Since
the scotogenic dark matter is leptophilic, indirect searches appear as an especially interesting possibility.
Here we propose a simple variation of
the  model with a very distinct phenomenology. The 
scotogenic fermion singlets are naturally grouped into pseudo-Dirac pairs of 
mass of $0.1$--$1$ TeV. We show that the lightest one constitutes a dark matter candidate that near threshold 
annihilates with a 90\%  branching ratio 
into neutrino pairs. 
The model gives the observed relic 
abundance consistently with the bounds from direct searches and with all neutrino and
charged lepton data. We also show that, for a sub-MeV dark matter particle, the model suggests a
scenario that could address the lithium problem.

\end{abstract}

\vfill
\eject

\section{Introduction: scotogenic dark matter} 
A neutrino mass $m_\nu\approx 0.05$ eV implies a 
Wilson coefficient $c_\nu/\Lambda\approx (10^{15}\,{\rm GeV})^{-1}$ in the Weinberg operator, suggesting 
that the UV completion of the neutrino sector involves physics at extremely
high energies. We have, however, a compelling reason to hope for new physics at a more accessible scale: the dark matter (DM) of the universe. Although one may argue that there are viable DM candidates of literally {\it any} mass,  the WIMP paradigm has provided the simplest and most popular framework. A GeV--TeV stable particle produced thermally in the early universe naturally implies the relic abundance deduced from observations, while its {\it invisibility} at the LHC and in direct searches favors models with stronger couplings to leptons than to quarks.
Scenarios where the DM also completes the neutrino sector appear then as an interesting possibility.

\begin{figure}[b]
\begin{center}
\hspace{-0.4cm}
\includegraphics[scale=0.86]{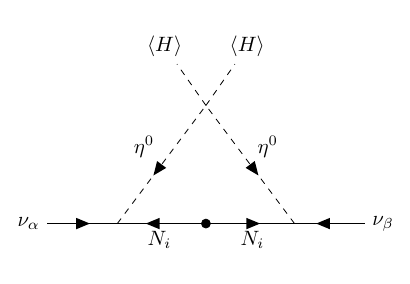}\hspace{0.2cm}
\includegraphics[scale=0.86]{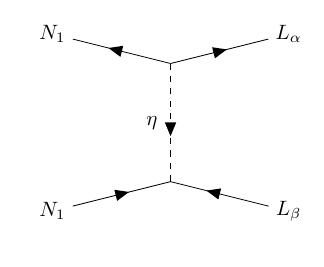}\hspace{0.cm}
\includegraphics[scale=0.86]{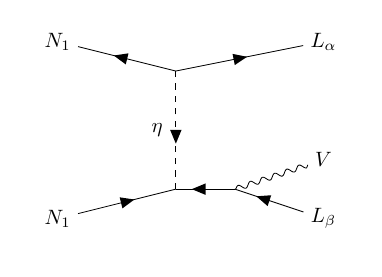}
\end{center}
\vspace{-0.5cm}
\caption{Diagrams generating neutrino masses and DM annihilation. 
\label{f1}}
\end{figure}

In this context, the scotogenic model \cite{Ma:2006km}, 
also referred to as leptophilic DM model or radiative seesaw model, has 
been extensively studied in the literature. The model is remarkable both because of 
its simplicity and its ability to explain  neutrino masses and the DM relic abundance 
consistently with the lack of signals in direct, indirect and collider searches 
\cite{Schmidt:2012yg,Garny:2015,Ibarra:2016dlb,Tang:2017,Beniwal:2020hjc,deBoer:2021pon,DeRomeri:2022cem, Babu:2025czb,Roy:2025moo}.
It introduces a heavy (Majorana) fermion singlet $N_i$ per light neutrino family and a second Higgs doublet 
$\eta=(\eta^+ \; \eta^0)$, both fields odd under an exact $Z_2$ matter parity that guarantees the stability of the lightest particle in this sector. In the basis of  $N_i$ and charged lepton ($\ell_\alpha$) mass eigenstates, the new interactions consistent with the symmetries read (in 2-component spinor notation\footnote{($\ell$, $\ell^c$, $N\dots$) are {\it left} and their conjugate ($\bar \ell$, $\bar \ell^c$, $\bar N\dots$) {\it right} spinors. We then define with primes Dirac $\ell'^T = (\ell \; \bar \ell^c)$ and Majorana 
$N'^T = (N \;\bar N)$ four-spinors, and $\eta^+ \ell N + \eta^- \bar \ell \bar N= \eta^+ \,\overline N' P_L \ell' +
\eta^- \overline \ell' P_R N' $.}, Lorentz and gauge indices contracted)
\beq
-{\cal L}\supset \sum_{i,\alpha} \left( {1\over 2} M_i \,N_i N_i + y_{\alpha i}\, \eta \,L_\alpha N_i + {\rm h.c.}\right) + V(H,\eta)\,,
\label{eq1}
\eeq
with 
\beqa
V&=&-\,\mu_1^2 \,H^\dagger H +\mu_2^2 \,\eta^\dagger \eta + \lambda_1 \,( H^\dagger H)^2+ 
\lambda_2 \,( \eta^\dagger \eta)^2 \;+\cr
&&\lambda_3 \,( H^\dagger H)\,( \eta^\dagger \eta) +
 \lambda_4 \,( H^\dagger \eta)\,( \eta^\dagger H) + 
 \displaystyle {\lambda_5\over 2}\left[ ( H^\dagger \eta)^2 + {\rm h.c.}\right] ,
\eeqa
where a possible complex phase in $\lambda_5$ is absorbed in a redefinition of $\eta$.
\begin{table}[t!]
\begin{center}
\begin{tabular}{|c|ccc|cc|}
\hline
 $\psi$ & \hspace{0.1cm} $L_\alpha $ \hspace{0.1cm}& \hspace{0.1cm} $e^c_\alpha$  \hspace{0.1cm} & 
 \hspace{0.1cm} $N_i$  \hspace{0.1cm} 
 & $H$  & \hspace{0.1cm} $\eta$  \hspace{0.1cm} \\ [0.4ex]
\hline 
$L(\psi)$ & $+1$ &
$-1$ & $0$ & $0$ & $-1$ \\ [0.4ex]
\hline
\end{tabular} 
\caption{Lepton number in the original scotogenic model, 
with $L_\alpha = (\nu\; \ell)_\alpha$, $H=(h^+\; h^0)$ and $\eta=(\eta^+ \; \eta^0)$.}
\label{L}
\end{center}
\end{table}
The lepton number assignment for the particles in this Lagrangian is given in Table \ref{L}, implying that only the 
term proportional to $\lambda_5$ breaks $L$. Therefore, the coefficient in the Weinberg operator must be proportional  to this parameter, that can consistently be taken much smaller than the rest of couplings. In particular, 
the loop diagram in Fig.~\ref{f1} defines a radiative seesaw mechanism for the neutrino masses \cite{Farzan:2012ev}. When the lightest $Z_2$-odd state is the fermion singlet $N_1$, then a relic density 
$\Omega_{N_1} h^2\approx 0.11$ for $M_1= 0.1$--$1$ TeV typically 
requires Yukawa couplings $y_1 = \left( \sum_\alpha |y_{\alpha 1} |^2\right)^{1/2} \gsim 0.1$, whereas neutrino masses  $m_\nu \approx 0.05$ eV are obtained for 
$\lambda_5\approx 10^{-9}$.

In addition to co-annihilations ({\it e.g.}, $N_1 \,\eta^+ \to \gamma \,e^+$),  the dominant annihilation mode of $N_1$ in the early universe
is $N_1 N_1 \to \ell^-_{\alpha} \ell^+_{\beta},\nu_\alpha \bar \nu_\beta$, that goes through the diagram in Fig.~\ref{f1} plus the corresponding $u$-channel 
contribution \cite{Ahriche:2021}. In the non-relativistic regime, however, if we neglect lepton masses the cross section is $p$-wave suppressed: the initial state must be antisymmetric under exchange, so an orbital $\ell=0$ implies $j=0$ and thus a chirality flip in the final (anti)lepton.
Expanding the cross section
in powers of the relative DM velocity, $\sigma \,v_{\rm rel} = a + b \,v_{\rm rel}^2$, we have
\cite{Schmidt:2012yg,Ibarra:2016dlb}
\beq
a=0\,;\hspace{0.5cm} b={M_1^2 \,y_{1}^4\over 48 \pi}
\left[ 
{M_1^4 + m_{\eta^+}^4\over (M_1^2 + m_{\eta^+}^2)^4} + 
{M_1^4 + m_{\eta^0}^4\over (M_1^2 + m_{\eta^0}^2)^4} 
\right].
\eeq
As a consequence, the dominant annihilation channel in astrophysical environments like the
Sun or the Galactic Center, where $v_{\rm rel}\approx \sqrt{3T/M_1}\ll 1$, is $N_1 N_1 \to L_{\alpha} 
\overline L_{\beta} V$ 
with $V=\gamma, W, Z$ \cite{Bell:2008,Garny:2011,Bell:2012rg,Fukushima:2012sp}  (see Fig.~\ref{f1}).

As already mentioned, this scotogenic model has attracted a lot of attention 
during the past 20 years. Despite that, we think that an interesting possibility may have been overlooked in previous literature. Here we would like to propose a simple variation of the model that includes 
the same particle content but a different definition of lepton number, implying a very distinct
phenomenology.

\section{The model} 
In the original scotogenic model the singlets $N_i$ have zero lepton number, and the inert doublet $\eta$
is a scalar {\it antilepton}. The Yukawa couplings $y_{\alpha i}$ in Eq.~(\ref{eq1}) respect lepton number and are then
unsuppressed, while the only coupling breaking $L$ is $\lambda_5\approx 10^{-9}$. Instead, 
we will assume  the presence of pairs $(N_i,N^c_i)$ of fermion singlets with lepton number $+1$ and $-1$, respectively, together with a different 
assignment to $\eta$, $L(\eta) = 0$. In the limit of exact lepton number, a Majorana
mass for the singlets and an interaction $\eta \,L_\alpha N_i$ are forbidden, whereas a Dirac mass, 
a $\lambda_5$ term in the scalar potential, and the Yukawa interaction 
$\eta \,L_\alpha N_i^c$ are all allowed. Let us consider the model with just one
$(N,N^c)$ pair. Including the terms that break $L$, the Lagrangian reads
\beq
-{\cal L}\supset \left( M \,N N^c + {\mu\over 2} \, NN + {\mu'\over 2} \, N^cN^c + 
\sum_\alpha \left(y_{\alpha}\, \eta \,L_\alpha N^c + 
\epsilon_{\alpha}\, \eta \,L_\alpha N \right) + {\rm h.c.}\right) + V(H,\eta)\,,
\label{eq4}
\eeq
where  $\epsilon_\alpha\ll y_\alpha$ and $\mu,\mu'\ll M$. 
\begin{table}[t!]
\begin{center}
\begin{tabular}{|c|cccc|cc|}
\hline
 $\psi$ & \hspace{0.1cm} $L_\alpha $ \hspace{0.1cm}& \hspace{0.1cm} $e^c_\alpha$  \hspace{0.1cm} & 
 \hspace{0.1cm} $N_i$  \hspace{0.1cm} &\hspace{0.1cm} $N^c_i$  \hspace{0.1cm} 
 & $H$  & \hspace{0.1cm} $\eta$  \hspace{0.1cm} \\ [0.4ex]
\hline 
$L(\psi)$ & $+1$ &
$-1$ & $+1$ & $-1$ & $0$ & $0$ \\ [0.4ex]
\hline
\end{tabular} 
\caption{New lepton number assignment.}
\label{LP}
\end{center}
\end{table}

\subsection{$L$-conserving limit}

It is instructive to first consider the
case with unbroken lepton number: $\epsilon_\alpha=0=\mu,\mu'$ and the two fermion singlets defining a Dirac fermion $N'$ of mass $M$ whose left and right handed components are $N$ and $\bar N^c$, respectively.
\begin{itemize}

\item
The first observation is that, obviously,  the standard neutrinos $\nu_\alpha$ stay massless as long as $L$ is unbroken.
\begin{figure}[t]
\begin{center}
\hspace{-0.4cm}
\includegraphics[scale=0.9]{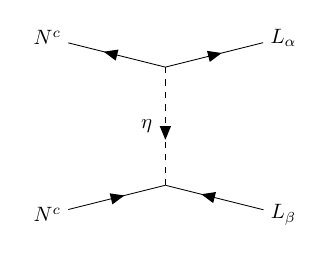}\hspace{0.5cm}
\includegraphics[scale=0.91]{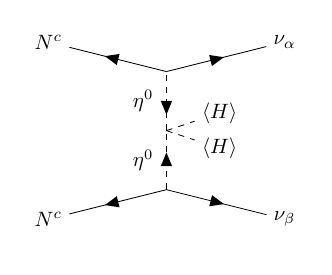}
\end{center}
\vspace{-0.5cm}
\caption{Annihilation channels $N' \bar N'\to \nu_\alpha \bar \nu_\beta, \ell_\alpha\bar\ell_\beta$ (left) and $N' N'\to \nu_\alpha \nu_\beta$ (right).
\label{f2}}
\end{figure}

\item
The second one, that two Dirac fermions $N'$ can annihilate in $s$ wave into lepton-antilepton and neutrino-neutrino pairs. In contrast with the Majorana case \cite{Hikasa:2024}, the initial state in $N' \bar N'\to L'_\alpha \bar L'_\beta$ (see Fig.~\ref{f2}-left) is not required to be antisymmetric. This first process is described by the
operator $(\overline{N'} \gamma^{\mu} P_R N')(\overline{L'}_\alpha \gamma_{\mu} P_L L'_\beta)$ 
\cite{Fuentes-Martin:2022jrf}, with 
${N'}^{T}=(N\; \bar N^c)$. As for 
$N' N'\to \nu_\alpha \nu_\beta$ (in Fig.~\ref{f2}-right, $u$ channel omitted), it is now an $L$-conserving and unsuppressed process where
both the initial and the final fermion pairs are in $s$ wave with total spin zero. 
Notice that, since the channel 
$N' N'\to \ell_\alpha \ell_\beta$ is forbidden by charge conservation, these neutrinos come without the corresponding charged leptons. 
In the usual 4-component spinor notation the process is described by the dim-8 operator 
$(H^\dagger \overline L'_\alpha P_R N')(H^\dagger \overline L'_\beta P_R N')$.
The  insertion in the 
$\eta^0$ line reflects the mass difference between ${\rm Re} \,\eta^0$ and ${\rm Im}\,\eta^0$,
\beq
m_R^2 - m_I^2 = \lambda_5 \,v^2\,,
\eeq
whose contributions would exactly 
cancel if $\lambda_5\to 0$. This would be the case in the usual scotogenic model when $M_1\approx M_2$ \cite{Schmidt:2012yg}. 
Notice also that $N^c$ couples to a single linear
combination of the three neutrinos  $\nu_\alpha$, with a coupling $y=\sqrt{|y_e|^2+|y_\mu|^2+|y_\tau|^2}$.
The $s$-wave contribution to the total annihilation cross section of the Dirac singlet is then\footnote{We have evaluated vertices and amplitudes using {\tt SARAH}~\cite{Staub:2013tta,Vicente:2015cka},
{\tt FeynArts}~\cite{Hahn:2000kx} and {\tt FeynCalc}~\cite{Mertig:1990an,Shtabovenko:2016sxi}.}
\beqa
a_{N'N'\to \nu\nu}&=&{M^2 \,y^4\over 256 \pi}
\left( {1\over M^2 + m_I^2} - {1\over M^2 + m_R^2} \right)^2 =a_{\bar N'\bar N'\to \bar\nu\bar \nu}\,;\cr
a_{N'\bar N'\to L \bar L}&=&{M^2 \,y^4\over 128 \pi}  \left( {1\over M^2 + m_I^2} + {1\over M^2 + m_R^2} \right)^2
+  \left( {2\over M^2 + m_{\eta^+}^2} \right)^2.
\label{aa}
\eeqa

\item
The final observation is that flavor changing processes like $\mu\to e \gamma$ and $\tau\to e\gamma$ will be 
unsuppressed unless the electron Yukawa coupling to $N^c$ is very small. This type of {\it texture} zeros
could reflect a flavor symmetry and are necessary in any scotogenic model. 

\end{itemize}

\subsection{$L$-breaking effects}

It is now straightforward to find the effects of the $L$-breaking terms in Eq.~(\ref{eq1}) on this Dirac model.

\begin{itemize}

\item
{\bf Spectrum.--}
The two components in the Dirac field {\it separate} into two Majorana fermions $N_1$, $N_{1'}$ of mass $M_1$  and 
$M_1+\Delta M_1$, respectively. Let us be more specific and consider the case 
\beq
-{\cal L}\supset{1\over 2}
\begin{pmatrix} N & N^c \end{pmatrix}
\begin{pmatrix} -\mu & M  \\ M & -\mu \end{pmatrix}
\begin{pmatrix} N \\ N^c \end{pmatrix},
\eeq
where, for simplicity, we have set $\mu'=\mu$ and have taken a negative Majorana mass coefficient ($\mu>0$). 
The mass eigenstates are obtained after the 
unitary transformation 
\beq
\begin{pmatrix} N_1 \\ N_{1'} \end{pmatrix} = {1\over \sqrt{2}}
\begin{pmatrix} 1 & 1 \\ i  & -i  \end{pmatrix}
\begin{pmatrix} N \\ N^c \end{pmatrix},
\eeq
and the eigenvalues are $M_1=M-\mu$ and $M_{1'}=M+\mu$, with
$\Delta M_1=2\mu\lsim 1$ MeV. 

\item 
{\bf Neutrino masses and mixings.--}
Both (quasi-Dirac) Majoranas  inherit from $N^c$ unsuppressed Yukawa couplings $y_{\alpha i}$ ($i=1,1'$) with the three lepton families:
\beq
-{\cal L}\supset \eta 
\begin{pmatrix} L_e & L_\mu &L_\tau \end{pmatrix}
\begin{pmatrix} \epsilon_e & y_e \\ \epsilon_\mu & y_\mu \\ \epsilon_\tau & y_\tau\end{pmatrix}
\begin{pmatrix} N \\ N^c \end{pmatrix} + {\rm h.c.} \hspace{3cm} \nonumber
\eeq
\beq
\hspace{1.cm}
\approx \eta 
\begin{pmatrix} L_e & L_\mu &L_\tau \end{pmatrix} {1\over \sqrt{2}}
\begin{pmatrix} \epsilon_e & -i \epsilon_e \\ y_\mu+\epsilon_\mu & 
i(y_\mu-\epsilon_\mu )\\ 
y_\tau+\epsilon_\tau  & i(y_\tau-\epsilon_\tau )\end{pmatrix}
\begin{pmatrix} N_1 \\ N_{1'} \end{pmatrix} +{\rm h.c.},
\eeq
where we have set $y_e\approx 0$ to suppress 
FCNCs involving the electron. The mass matrix generated at one loop reads then
\beq
(m_\nu)_{\alpha \beta} \approx {M_1 \over 16\pi^2 } \left(  (Y\cdot Y^T)_{\alpha \beta}+ {\Delta M_1  \over  M_1}\,
Y_ {\alpha 1'} Y_ {\beta 1'} \right)\left( {m_R^2\over m_R^2 - M_1^2}
\ln {m_R^2\over  M_1^2}  - {m_I^2\over m_I^2 - M_1^2} \ln {m_I^2\over  M_1^ 2} \right),
\eeq
where the dominant elements 
in  $(Y\cdot Y^T)_{\alpha \beta}+ {\Delta M_1  \over  M_1}\,
Y_ {\alpha 1'} Y_ {\beta 1'}$  are
\beq
\begin{pmatrix}
0 & \epsilon_e y_\mu & \epsilon_e y_\tau \\
\epsilon_e y_\mu & (2\epsilon_\mu -\epsilon_My_\mu )\,y_\mu & \epsilon_\tau y_\mu + \epsilon_\mu y_\tau -2 \epsilon_My_\tau y_\mu \\
\epsilon_e y_\tau &  \epsilon_\tau y_\mu + \epsilon_\mu y_\tau -2 \epsilon_My_\tau y_\mu
&  (2\epsilon_\tau-\epsilon_M y_\tau)\, y_\tau 
\end{pmatrix},
\label{mnu}
\eeq 
with $\epsilon_M\equiv \Delta M_1/(2M_1)$.
The matrix above defines a radiative inverse-seesaw mechanism \cite{Ma:2009gu,Nga:2025rgz} for the neutrino masses. It 
has a zero eigenvalue ($m_1=0$) but can fit the central value of the two neutrino mass differences and the two larger mixings. For $\theta_{13}$ it implies
\beq
m_2 \cos^2\theta_{13} \sin^2\theta_{12}-m_3 \sin^2\theta_{13}=0
\eeq
and thus $\sin^2\theta_{13}=(4.55\pm 0.20)\times 10^{-2}$, versus the current best fit value
$\sin^2\theta_{13}=(2.16\pm 0.06)\times 10^{-2}$ \cite{Esteban:2024}. This {\it prediction} is a consequence of taking $y_e\approx 0$ (to prevent FCNCs involving the electron) together with
$y_{\mu,\tau}=O( 0.1)$, and its correction requires that the model includes at least an additional $(N,N^c)$ pair (see next section).

\item 
{\bf Annihilations.--}
The dominant channels of annihilation and co-annihilation between the two quasi-degenerate fermion
singlets $(N_1,N_{1'})$ can be obtained from our results for the Dirac fermion $N'$ in the previous subsection.
In particular, for the $s$-wave processes we have
\beq
a_{N_1N_1},a_{N_{1'}N_{1'}}\simeq a_{N'N'}\,;
\qquad
a_{N_1N_{1'}}\simeq a_{N'\bar N'}\,.
\eeq
In environments with no $N_{1'}$ states co-annihilations will be absent and the only two-body
$s$-wave channel processes are
\beq
N_1N_1\to \nu_\alpha\nu_\beta ,\, \bar\nu_\alpha\bar\nu_\beta\,.
\eeq
In this case electroweak bremsstrahlung 
$N_1N_1\to L_\alpha\bar L_\beta V$ will typically account for 10\% of all the annihilations.

\begin{figure}[b]
\begin{center}
\hspace{-0.4cm}
\includegraphics[scale=0.88]{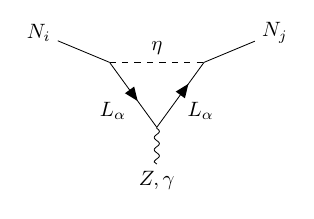}\hspace{0.2cm}
\includegraphics[scale=0.88]{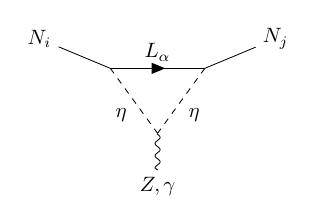}\hspace{0.1cm}
\includegraphics[scale=0.88]{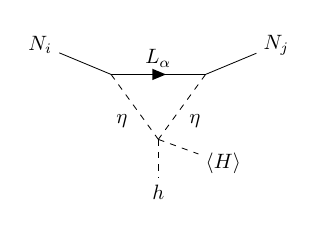}
\end{center}
\vspace{-0.5cm}
\caption{Diagrams generating transitions and elastic collisions through the $\gamma$, $Z$ and Higgs portals. 
\label{f3}}
\end{figure}
\item 
{\bf Decays, inverse decays and other $N_1 \to N_{1'}$ transitions.--}
The one-loop diagrams in Fig.~\ref{f3} generate a transition magnetic moment 
[at this point we switch to 4-spinors and omit the prime, {\it e.g.}, ${N'_1}^T=(N_1\;\bar N_1)={N_1}^T$ or
${\ell'}^T=(\ell\;\bar \ell^c)={\ell}^T$]
\beq
{\cal L}_{eff} \supset {i\over 2} \,\mu_{1'1} \overline N_{1'}\,\sigma^{\mu\nu} N_1 F_{\mu \nu}\,,
\eeq
with \cite{Schmidt:2012yg}
\beq
\mu_{11'}\approx -{e \, y^2\over 32\pi^2}\, {M\over m_{\eta^+}^2}\, F\!\left({M^2\over m_{\eta^+}^2}\right);
\hspace{0.5cm}
F(x)={-x-\ln (1-x)\over x^2}\,.
\eeq
This implies that the heavier Majorana decays $N_{1'}\to N_1 \gamma$ with a partial width  
\beq
\Gamma_\gamma\approx {\mu_{11'}^2\over \pi}\, (\Delta M_1)^3\,.
\eeq
$N_{1'}$ can also decay into neutrinos: 
\beq
\Gamma_{\nu}\approx 
{y^4\over 960\pi^3} \left[ \left({1\over m_I^2-M_1^2} - {1\over m_R^2-M_1^2}\right)^ 2 
+\left( {1\over m_I^2-M_1^2} + {1\over m_R^2-M_1^2}\right)^2 \right]
(\Delta M_1)^5\,,
\eeq
with the first term from
$N_{1'}\to N_1 \nu \nu , N_1 \bar \nu \bar \nu$ and the  second one from $N_{1'}\to N_1 \nu \bar \nu$.
Despite being a tree-level process, this channel is mediated by a dim-6 operator (versus the dim-5 electromagnetic 
transition) and   is negligible for $\Delta M_1\approx$ keV--MeV.

In astrophysical environments with a hot plasma at $T\gsim \Delta M_1$, $N_{1'}$ may be produced via inverse decays $N_1 \gamma \to N_{1'}$ and in collisions with neutrinos and electrons. To estimate the relevance of the first process we take a Boltzmann
photon distribution and a non-relativistic $N_1$, 
\beq
{{\rm d} n_\gamma \over {\rm d} E}\approx{ E^2\over \pi^2}\, e^{-E/T}\,; \hspace{0.5cm}
\sigma_\gamma \approx {\pi^2\over E^2}\, \Gamma_\gamma\,\delta (E-\Delta M_1)\,.
\eeq
The rate of $N_1 \gamma\to N_{1'}$ transitions is then
\beq
\Gamma_\gamma^{\rm tr}\approx \Gamma_\gamma \,e^{-\Delta M_1/T}\,.
\eeq

If $y_e\not = 0$ 
the transition can also go through $N_1 \nu_e \to N_{1'} \nu, N_{1'} \bar \nu$ and
$N_1 e \to N_{1'} e$. In the limit $M_1\gg E>\Delta M_1$ the inclusive cross section  is just
\beq
\sigma_{\nu_e} \approx {|y_e|^2y^2 \left( E-\Delta M_1 \right)^2\over 64\pi}
\left(
{1\over \left(m_I^2-M_1^2\right)^2}  + {1\over \left(m_R^2-M_1^2\right)^2} \right),
\eeq
and in environments with a ${\rm d} n_{\nu}/{\rm d} E$ number density of $\nu_e$ the rate
becomes
\beq
\Gamma_\nu^{\rm tr}\approx 
{|y_e|^2y^2 \over 64\pi}
\left(
{1\over \left(m_I^2-M_1^2\right)^2}  + {1\over \left(m_R^2-M_1^2\right)^2} \right)
\int_{\Delta M_1}^\infty {\rm d}E\, {{\rm d} n_{\nu}\over {\rm d} E} \left( E-\Delta M_1 \right)^2.
\eeq
For example, $pp$ neutrinos at the solar core reach 
${\rm d} n_{\nu}/{\rm d} E\approx 10^4$ cm$^{-3}$ keV$^{-1}$ at energies $E\le 420$ keV. If 
$\Delta M_1 \le 10$ keV, however, the number of thermal electrons with kinetic energy $E-m_e>\Delta M_1$ able to mediate the transition
through $N_1 e \to N_{1'} e$ is much larger than the number of neutrinos, and
\beq
\sigma_e\approx {|y_e|^4 \,m_e^2\over 64\pi \,( m_{\eta^+}^2-M_1^2)^2} 
\sqrt{1-{\Delta M_1\over E-m_e}}\,.
\eeq
In a point inside the Sun with mass density $\rho$ and temperature $T$ we have
\beq
{{\rm d} n_e \over {\rm d} E}\approx {3\rho \over m_p} \sqrt{4 (E-m_e)\over \pi T^3} e^{-(E-m_e)/T},
\eeq
implying
\beq
\Gamma_e^{\rm tr}\approx 
\int_{m_e+\Delta M_1}^\infty {\rm d}E\, {{\rm d} n_{e}\over {\rm d} E} \,v_e\,\sigma_e
\approx {3\rho\over m_p}\,{|y_e|^4 \,m_e^{3/2} \,T^{1/2}\over 32\sqrt{2}\,\pi^{3/2} \,( m_{\eta^+}^2-M_1^2)^2}
\,e^{-\Delta M_1/T},
\eeq
where we have taken $(\Delta M_1, T)\ll m_e$. From all these expressions it follows, however, that in the Sun the
$N_1\leftrightarrow N_{1'}$ transitions involving
electrons and neutrinos are negligible respect to $N_{1'} \leftrightarrow N_1 \gamma$.

\item 
{\bf Interaction with matter.--}
Depending on their relative velocity $\vrel$, the scattering of a DM particle $N_{1}$ off a nucleus $A$ can be elastic ($N_1 A \to N_1 A$) or inelastic ($N_1 A \to N_{1'} A$). The inelastic collision is kinematically allowed only if \cite{deBoer:2021pon} 
\beq
  \Delta M_1 < \frac{1}{2}\,\mred\,\vrel^2 \, ,
\eeq
with $\mred = M_1\, m_A/(M_1+m_A)$.
As shown in Fig.~\ref{f3},
these processes may be mediated by the photon, the $Z$ boson or the Higgs boson. 
In Table \ref{eff} we summarize the different effective operators involved (we have used
the {\tt Matchete} package \cite{Fuentes-Martin:2022jrf}).
\begin{table}[b]
  \centering
  \begin{tabular}{lccc}
    \hline
    Operator & \hspace{0.5cm}elastic\hspace{0.5cm} & inelastic & \hspace{0.5cm}mediator\hspace{0.5cm} \\
    \hline
    $(\overline N_i N_1)\,(\bar q q)$
      & $S^{h}_{11}$ & $S^{h}_{1'1}$ & $h$ \\

    $(\overline N_i\gamma^\mu\gamma_5 N_1)\,
      \bigl(\overline q\gamma_\mu (V_q^Z-A_q^Z\gamma_5) q\bigr)$
      & $A^{Z}_{11}$ & $A^{Z}_{1'1}$ & $Z$ \\

    $(\overline N_{1'}\gamma^\mu N_1)\,
      \bigl(\overline q\gamma_\mu (V_q^Z-A_q^Z\gamma_5) q\bigr)$
      & $0$ & $i\,V^{Z}_{1'1}$ & $Z$ \\

    $\overline N_i\gamma^\mu\gamma_5 N_1\,\partial^\nu F_{\mu\nu}$
      & ${\cal A}_{11}$ & ${\cal A}_{1'1}$ & $\gamma$  \\

    $\overline N_{1'}\gamma^\mu N_1\,\partial^\nu F_{\mu\nu}$
      & $0$ & $i \,a_{1'1}$ & $\gamma$ \\

    $\overline N_{1'}\sigma^{\mu\nu}N_1\,F_{\mu\nu}$
      & $0$ & $i\,\mu_{1'1}/2$ & $\gamma$ \\

    $\overline N_{1'}\gamma^\mu N_1\,A_\mu$
      & $0$ & $i\,c_{1'1}$ & $\gamma$  \\
    \hline
  \end{tabular}
  \caption{Operators relevant for $N_1$--nucleus scattering ($i=1,1'$).  $V_q^Z$ and $A_q^Z$ denote the vector and axial-vector quark couplings to the $Z$ boson.}
\label{eff}
\end{table}

For elastic scattering, $S_{11}^h$ and $A_{11}^Z$ encode the one-loop effective interactions mediated by $h$ and $Z$, respectively,  
while ${\cal A}_{11}$ denotes the electromagnetic anapole moment
\cite{Ho:2012bg,Kopp:2014tsa,Herrero-Garcia:2018koq,Ibarra:2022trb}. 
In the non-relativistic limit \cite{Fitzpatrick:2013,Anand:2014, Gresham:2014, DelNobile:2018,Catena:2015uha}, 
$S_{11}^h$ implies a spin independent
$N_1$--nucleon contact interaction $V(x)=c_1^N\,\delta(\vec x)$ with total 
cross section $\sigma_{SI}^N=(\mu_{1N}^2/\pi)\,|c_1^N|^2$,  where \cite{Ibarra:2016dlb,Babu:2025czb} 
$\mu_{1N} \equiv M_1\, m_N/(M_1+m_N)$, $m_N=0.94$ GeV,  
\beq
c_1^N={f_N m_N \, y^2\over 16\pi^2 \,m_h^2 M_1}\left( \lambda_3 \,{\cal G}_1(M_1^2/m_{\eta^+}^2) 
+{\lambda_3 + \lambda_4\over 4}\left( {\cal G}_1(M_1^2/m_I^2) + {\cal G}_1(M_1^2/m_R^2) \right)  
\right),
\eeq
$ {\cal G}_1(x)=\left[ x +(1-x)\ln (1-x) \right]/x$ and $f_N=0.3$ (we assume isospin-conserving scalar matrix elements).

The term proportional to $ A^{Z}_{11} A_q^Z$ yields a spin-dependent interaction 
$V(\vec x)=c_4^N\,\vec S_{N_1}\cdot \vec S_N\, \delta(\vec x)$ with total cross section  
$\sigma_{SD}^N=(3\mu_{1N}^2/(16\pi))\,|c_4^N|^2$. We obtain \cite{Ibarra:2016dlb, Babu:2025czb}
\beq
c_4^N=- {y^2\,a_N\,g\over 16\pi^2 \,m_Z^2 \,c_W}\left(  \left( 1- 2 s_W^2 \right) \,{\cal G}_2(M_1^2/m_{\eta^+}^2) 
-  \left( 1+\ln {m_R^2\over m_I^2} \right) {\cal G}_2(M_1^2/m_I^2) \right),
\eeq
with $a_p=0.677$, $a_n=-0.592$ \cite{HERMES:2006jyl} and $ {\cal G}_2(x)= -1 + 2 \left[ x+(1-x)\ln (1-x) \right]/x^2$.
The anapole interaction, in turn, leads to a velocity-suppressed scattering \cite{Ibarra:2016dlb,Anand:2014} that will be subleading in processes
like direct detection or DM capture by the Sun.

Finally, in the same non-relativistic limit the elastic scattering $N_1 e \to N_1 e$ has a total cross section  
\beq
\sigma^e_{\rm el} \approx {3\,|y_e|^4\,\mu_{Ne}^2\over 64\pi \,( m_{\eta^+}^2-M_1^2 )^2}\,,
\label{elel}
\eeq
with $\mu_{Ne}\approx m_e$. In contrast with quark scattering, this leptonic interaction arises already 
at the tree level, through the exchange of the charged component in the inert scalar doublet.

As for the inelastic processes, the dipole  electromagnetic transition mediating $N_{1'}\to N_1 \gamma$ and 
$N_1 e \to N_{1'} e$ has been already discussed in the previous paragraph. The inelastic upscattering
$N_1 A \to N_{1'} A$ includes processes analogous to the elastic ones. Generically, these processes imply
smaller nuclear recoils than the elastic ones and will introduce subleading effects in direct detection experiments. They can, however, play an important role in the 
DM capture by an astrophysical object: despite this smaller recoil in the initial collision of $N_1$, 
the final decay $N_{1'}\to N_{1}\gamma$ adds $E_\gamma\approx \Delta M_1$ to the total energy lost by the projectile. The four
transitions in Table \ref{eff} associated to bilinears that vanish for a single Majorana fermion 
($\overline N_1 \gamma^\mu N_1=0=\overline N_1 \sigma_{\mu\nu} N_1$)
can also mediate inelastic spin dependent and spin independent interactions with a nucleus of atomic number $Z$, spin $J_A$ and magnetic moment $\mu_A$ (see \cite{Schmidt:2012yg} for a detailed evaluation of the cross sections).

\end{itemize}

\section{Dark matter scenarios} 
Some phenomenological aspects of the DM model proposed in the previous section depend critically on the
mass difference $\Delta M_1=M_{1'}-M_1$. Other observables, however, are insensitive to the particular value in the
keV--MeV range of this parameter. In particular, the relic abundance is obtained at cosmological temperatures
$T_{\rm f.o.}\approx M_1/20\gg \Delta M_1$. At these temperatures the abundances of $N_{1}$ and $N_{1'}$ are very similar: the two Majoranas evolve 
as if they were a single Dirac fermion of mass $M$ and the final relic abundance could be estimated using 
the annihilation cross section in Eq.~(\ref{aa}). We will, however, obtain a more precise estimate of
$\Omega_{1,1'}h^2$ with {\tt micrOMEGAs}
\cite{Belanger:2006is,Alguero:2022isa}, after implementing the model in {\tt SARAH}~\cite{Staub:2013tta} and generating the
{\tt CalcHEP} format files \cite{Belyaev:2012qa}. In our numerical computation we will 
keep the full velocity dependence of the cross sections and will include  all the relevant
coannihilations with $N_{1'}$ and with the neutral and charged components in $\eta$ \cite{Griest:1990kh}.

Although the parameter space of the model is very wide, to illustrate its main features it may be useful to use  benchmark points. We take three pairs of quasi-Dirac fermions
 $N_{i,i'}$ of masses $M_1=200$ GeV, $M_2=500$ GeV, $M_3=1$ TeV. We assume that the component $N^c_i$ in
 each pair couples only to one lepton family, while the $N_i$ component in each pair 
 couples to the three lepton families:
 \beq
-{\cal L}\supset
y_1 \,\eta L_e N^c_1+y_2 \,\eta L_\mu N^c_2+y_3 \,\eta L_\tau N^c_3 \hspace{2cm}\nonumber 
\eeq
\beq
\hspace{1.5cm}+\sum_\alpha \left( \epsilon_{\alpha 1} \,\eta L_\alpha N_1+ \epsilon_{\alpha 2} \,\eta L_\alpha N_2+ 
\epsilon_{\alpha 3} \,\eta L_\alpha N_3 \right)
+{\rm h.c.}
\label{benchmark}
\eeq
The relic abundance depends then only on the coupling $y_1$ and on the spectrum of scalar mediators. Taking
$y_1=0.4$, $m_{\eta^+}=225$ GeV, $m_I=211$ GeV and $\lambda_5=0.7$, which corresponds to $m_R=296$ GeV 
(the masses have been calculated at the one-loop level \cite{Goudelis:2013} with {\tt SPheno}~\cite{Porod:2011nf}),
we obtain $\Omega_{1,1'}h^2\approx 0.11$. Throughout our discussion we will keep these values as default and set also $y_{2,3}=y_1$. 

A second feature that is independent from the precise value of $\Delta M_1$ is the $N_1$ elastic cross section
off nucleons and electrons. Our benchmark gives
\beq
\sigma^e_{\rm el}=1.6\times 10^{-45}\;{\rm cm}^2\,;\;\;
\sigma^N_{\rm SD}=2.5\times 10^{-45}\;{\rm cm}^2\,;\;\;
\sigma^N_{\rm SI}=4.6\times 10^{-48}\;{\rm cm}^2\,.
\label{elcs}
\eeq
As expected in scotogenic models, these cross sections are well below the limits established by direct search experiments. For the scattering off nucleons, we notice that the model naturally accommodates a spin dependent interaction much larger than the spin independent one. As for the elastic cross section with the electron, 
Eq.~(\ref{elel}) shows that  it could be substantially larger in models where $m_{\eta^+}\to M_1$. 

Next we fix the couplings $\epsilon_{\alpha i}$ that reproduce the neutrino masses and mixings: $m_1=0.005$ eV, $m_2=0.01$ eV, $m_3=0.05$ eV, $\theta_{12}=0.587$, $\theta_{23}=0.819$, $\theta_{13}=0.147$, and $\delta=3\pi/2$.
As shown in Eq.~(\ref{mnu}), the value of these couplings depend on the particular value of $\Delta M_i$. For example, if $\Delta M_i=1$ keV for the three quasi-Dirac singlets we can take
\beq
\epsilon_{\alpha i}=10^{-10}\times
\begin{pmatrix}
12.4 & 0 & 0.18+0.18\,i \\
0.24-0.27\,i & 5.6-0.02\,i & 0 \\
0 & 0.89 & 2.0-0.02\,i
\end{pmatrix}.
\label{benchmark_epsilon}
\eeq
Larger values of $\Delta M_i$ require either to fine tune $\epsilon_{\alpha i}$ 
or smaller values of $\lambda_5$; notice that when 
$\lambda_5 \to 0$
we recover the usual scotogenic model, with $\Delta M_1\approx M_i$.
Our choice with only one quasi-Dirac fermion pair sizeably coupled to each lepton family 
suppresses FCNCs almost to zero regardless of the value of $\Delta M_i$. 
In particular \cite{Toma:2013zsa},
\beq
{\rm BR}(\mu \to e \,\gamma)={3\,\alpha_{EM} \,{\rm BR}(\mu \to e \,\nu_\mu \bar\nu_e)\over 64\pi \,G_F^2 \,m_{\eta^+}^4 } \left| y_1\,\epsilon_{\mu 1}\,F_2({M_1^2/ m_{\eta^+}^2}) + 
y_2\,\epsilon_{e 2}\,F_2({M_2^2/ m_{\eta^+}^2})
\right|^2,
\eeq 
with $F_2(z)=(1/6-z+z^2/2+z^3/3-z^2\ln z)/(1-z)^4$. We obtain ${\rm BR}(\mu \to e \,\gamma)\approx 10^{-26}$, with
similar branching ratios for $\tau \to e \,\gamma$ and $\tau \to \mu \,\gamma$.

Let us then sketch the different scenarios that are defined by the different values of 
$\Delta M_1$. Again, to be definite we will comment just on four values of $\Delta M_1$ that we consider representative.

\begin{itemize}

\item $\Delta M_1 =0.5\;{\rm keV}$

For the benchmark value of $y_1$ and $M_1$ assumed, 
the lifetime of $N_{1'}$ is 1 yr. In the early universe, this implies a decay before recombination,  when the photon temperature is around $100$ eV.
The injected radiation, of $E_\gamma=0.5$ keV, 
will not affect the abundance of primordial nuclei nor introduce CMB spectral distortions.

In astrophysical environments with $T\gsim \Delta M_1$, like the Sun's core or the hot plasma in the Galactic Center, 
decays and inverse decays
$N_{1'}\leftrightarrow N_1 \gamma$ will keep similar abundances of both species, and the DM will 
 be an effective Dirac particle. The dominant annihilation channels there would  be to monochromatic neutrinos and  electrons:
 $\nu_e\nu_e+ \bar\nu_e\bar\nu_e+ \nu_e\bar\nu_e$  (44\%)  and  $e^+e^-$ (55\%).

\item $\Delta M_1 =10\;{\rm keV}$

At these larger mass splitting $N_{1'}$ will not be thermally produced, but for $\vrel\approx 10^{-3}$ the 
$N_1\to N_{1'}$ transitions will be kinematically open 
in DM collisions with matter. As mentioned before, these inelastic processes will be subleading in 
direct search experiments, but they would contribute to the capture of DM by an astrophysical object. Notice that, while in  the elastic collision with a He nucleus the energy lost by the incoming $N_1$ is always below 
$2 m_{\rm He} \vrel^2\approx 8$ keV, in inelastic collisions 
($N_1 {\rm He} \to N_{1'}  {\rm He} \to N_1 \gamma  {\rm He}$) the final  photon adds 10 keV  to the energy lost by the DM particle. 

At this benchmark point, 
in any environment where $T<\Delta M_1$ the dominant annihilation  channel  is   
$\nu_e\nu_e+ \bar\nu_e\bar\nu_e$  (88\%), with annihilations to three body final states 
({\it e.g.}, $e^- \bar  \nu \, W^+$) accounting for the rest (12\%).

\item $\Delta M_1 =100\;{\rm keV}$

In this regime the $N_1$  to $N_{1'}$ transitions do not occur  thermally nor through collisions with  nuclei. In particular, the Sun will capture DM only via elastic scattering with nuclei and electrons. 
In our benchmark case the elastic cross sections [in Eq.~(\ref{elcs})] are very small and thus 
the capture rate weak; we estimate
\beq
C^e=5.6\times 10^{15}\,{\rm s}^{-1}\,;\;\;C^{\rm SI}=1.9\times 10^{17}\,{\rm s}^{-1}\,;\;\;
C^{\rm SD}=2.8\times 10^{17}\,{\rm s}^{-1}\,.
\eeq
The model, however, admits cases with much larger elastic cross sections. 
For example, if $M_1=60$ GeV, $y_1=0.38$, 
$m_I=78$ GeV, $m_R=437$ GeV, $m_{\eta^+}=308$ GeV and $\lambda_5=3$, we obtain $\Omega_{1,1'}h^2=0.11$ and 
\beq
\sigma^e_{\rm el}=2.4\times 10^{-48}\;{\rm cm}^2\,;\;\;
\sigma^N_{\rm SD}=3.2\times 10^{-42}\;{\rm cm}^2\,;\;\;
\sigma^N_{\rm SI}=3.9\times 10^{-48}\;{\rm cm}^2\,.
\label{elcs2}
\eeq
These spin-dependent and spin-independent cross sections are near the current bounds from direct searches
\cite{Aalbers:2024}, and they imply a much higher capture rate:
\beq
C^e=8.9\times 10^{13}\,{\rm s}^{-1}\,;\;\;C^{\rm SI}=7.7\times 10^{17}\,{\rm s}^{-1}\,;\;\;
C^{\rm SD}=3.5\times 10^{21}\,{\rm s}^{-1}\,.
\eeq
They induce a solar emission of  
$\Gamma_A = ({C/2})\tanh^2(t_\odot/ \tau)\approx 1.7\times 10^{21}$ neutrino pairs per second, or
$3.9\times 10^{11}$ monochromatic neutrinos of 60 GeV reaching the Earth per km$^2$ and yr. 

\item $\Delta M_1 =1.7\;{\rm MeV}$

This is an intriguing possibility, as the decay $N_{1'}\to N_1 \gamma$ in the early universe could affect
BBN \cite{Pospelov:2010hj,Kawasaki:2004qu}. More precisely, an injection of 1.58--2.22 MeV photons 
at $3\times 10^4\;{\rm s}\le t \le 10^6\;{\rm s}$ ($T= 1$--$6$ keV) would
break the $^7$Be  (${\rm ^7Be}+\gamma \to {\rm ^3He}+ {\rm ^4He}$) before it decays into $^7$Li,
while not affecting the primordial abundance of D or He.
In our framework it would require, first of all, a lighter DM particle, as the benchmark mass 
$M_1=200$ GeV together with the relatively large values
of $\Delta M_1$ and $\mu_{1'1}$ imply a too short lifetime, $\tau_{1'}<1$ s. For example, in a model with
$M_1=10$ GeV and $y_1=0.07$,  $m_I=13$ GeV, 
$m_{\eta^+}= 495$ GeV and $m_R=350$ GeV, the lifetime becomes 
$\tau_{1'}=3\times 10^4\;{\rm s}$. 

The problem is, however, that to break a significant fraction of  primordial $^7$Be  and solve
the Li problem we need a very large number density of decaying particles $N_{1'}$
\cite{Poulin:2015woa}, {\it i.e.}, a very
large final number density of DM particles. This implies that $M_1$ 
should be sub-MeV. Let us be a bit more specific and assume that
$M_1=60$ keV, $M_{1'}=3.8$ MeV and 
$\tau_{1'}=3\times 10^4$ s. The sequence of events
should go as follows. Both species are in 
thermal equilibrium with photons, electrons and neutrinos until $T\approx 4$ MeV, when the abundance of
the heavier partner goes down exponentially. Then $N_{1'}$ freezes out when 
$n_{1'}/n_\gamma \approx 7\times 10^{-7}$.
The lighter species $N_1$, still in thermal equilibrium, becomes  
non-relativistic during BBN, when its abundance drops 
exponentially until it decouples at $n_1/n_\gamma=5\times 10^{-5}$.
This cold DM component will dominate the final $N_1$ energy density. The second component appears
when $N_{1'}$ decays into $N_1\gamma$ 
at $T\approx 6$ keV. The 1.9 MeV photons deplete then most of 
the $^7$Be (our parameter $\zeta_{N_{1'}\to \gamma} = n_{1'}E_\gamma/n_\gamma=1.4\times 10^{-6}$
sits at the lower edge of the viable region identified in \cite{Poulin:2015woa}), whereas the ultra-relativistic
$N_1$ fermions, once they are stopped by the expansion, will constitute around $1.3\%$ of the total DM of the universe.

\end{itemize}

\section{Summary and discussion} 

The search for DM has been a priority in the field, specially after the LHC {\it completed} the SM 
with the discovery of 
the Higgs boson in 2012. Given the wide range of possibilities, it has been essential to identify
generic features that may be shared by many models and can then guide the
search. Within the WIMP paradigm, the scotogenic model has been very valuable, as it provides
a framework to understand a leptophilic candidate. The model may look
too simple or {\it ad hoc}, but the fact that it is based on symmetries brings technical consistency to  
a set up that is able to unify the neutrino and the DM sectors, indeed a remarkable feat.
One could expect that {\it any} TeV model where the DM 
interacts with baryons only at the loop level and explains the tiny size of neutrino masses 
shares a similar phenomenology. 

The simple variation of the model proposed here gives a complementary set
of features and defines an also consistent playground for quasi-Dirac DM. The variation includes
basically the same matter content as the usual scotogenic model, but with a different lepton
number assignment: instead of $L(N_i)=0$ and $L(\eta)=-1$,  we take 
$L(N_i)=+1$, $L(N^c_i)=-1$ and $L(\eta)=0$, with $(N_i, N^c_i)$ becoming a  pair
$(N_i,N_{i'})$ of Majorana fermions with a mass splitting $\Delta M_i\ll M_i$.
In the non-relativistic limit,
the DM annihilates predominantly into neutrino pairs: while in the original model 
$N_1 N_1\to \nu \nu$ is suppressed by lepton number
conservation or  $N_1 N_1\to \ell^+ \ell^-$  is reduced by a factor of $m_\ell^2/M_1^2$, 
in this set up the dominant annihilation channel is 
$N_1 N_1\to \nu \nu$ (the $p$-wave suppression in $N_1 N_1\to \ell^+ \ell^-$ persists). We believe
that this is the most striking feature of the model: it defines an UV complete framework where the WIMP 
may annihilate 90\% of the times into monochromatic
neutrinos. In contrast, the usual
scotogenic model with two Majorana singlets of similar mass, a case often considered
in the literature, implies that the annihilation of the lightest one near threshold is 
electroweak bremsstrahlung ({\it e.g.}, $N_1 N_1\to \ell^+ \ell^- Z$). 

Another interesting feature of the model is that it may behave very differently in astrophysical environments at
different temperature. We have shown that in a hot plasma at $T\ge \Delta M_1$ 
decays and inverse decays $N_{1'}\leftrightarrow N_1\,\gamma$  keep similar abundances of the two
species. In that case the DM becomes an effective Dirac particle and the dominant annihilation channels will also include monochromatic charged leptons,  $N_1 N_{1'}\to \ell^+ \ell^-$.

Monochromatic neutrinos can be searched at km$^3$ telescopes in contained $\nu_e$ events. While
the directionality in these events is very poor, the incident neutrino would deposit {\it all} its energy, and  the telescope becomes a calorimeter with a much better energy resolution than in $\nu_\mu$ events. In a particular DM model discussed before we have 
estimated a solar flux of $3.9\times 10^{11}$ neutrinos per km$^{2}$ and yr, all of them at
$E_\nu=60$ GeV. If produced with the electron flavor (the case we have assumed), 
55\% of them will reach the Earth with the same flavor.
For comparison, the total flux of $E_\nu \ge 10$ GeV neutrinos from CR showers on the solar surface 
 is 230 times smaller \cite{delaTorre:2025llo, Masip:2017gvw}.
The atmospheric flux of electron neutrinos with energy $50\; {\rm GeV}\le E_\nu\le 70\; {\rm GeV}$ within a $10$ degree circle around the Sun at zenith $\theta_z=-45^\circ$ is $9\times 10^{9}$ per km$^{2}$ and yr \cite{Lipari:1993hd}, suggesting that the model could be at the reach of KM3NeT \cite{Gozzini:2025ivt} and IceCube \cite{IceCube:2023ies}.
Generically, the model provides solar fluxes more accessible at telescopes for 
scalar masses $m_{\eta^+}$ closer to $M_1$ (that increase the capture rate
via elastic collisions with electrons  \cite{Nguyen:2025ygc}), 
lower values of 
$M_1$ (implying a more effective capture via spin dependent collisions with protons \cite{Nguyen:2026nhe}) and lower values also of 
$\Delta M_1$ (so that inelastic collisions with nuclei are above threshold).

We have finally discussed a possible scenario where the decay $N_{1'}\to N_1\,\gamma$ after BBN could break 
the primordial $^7$Be before it decays into $^7$Li.
It  would resemble the one proposed in \cite{Salvati:2016jng}, where a sterile MeV neutrino decays into a gamma plus a standard neutrino. In our model, the decay of $N_{1'}$ into the DM particle appears in a very narrow corner of the parameter space, where the general analysis in previous sections does not apply. Although it is then unclear whether the set up is flexible enough to actually accommodate it, the potential link of the lithium problem to the origin of DM is very suggestive and 
may deserve a dedicated analysis.

In summary, several decades of direct and indirect searches have constrained substantially the parameter space of a DM WIMP. However, as long as there are possibilities still open, it seems worth to identify them and explore them.

\vfill
\eject

\section*{Acknowledgments}
We would like to thank Javier Fuentes-Martin, Renato Fonseca and Manuel P\'erez-Victoria for discussions.
This work has been supported by the Spanish Ministry of Science, Innovation and Universities 
MICIU/AEI/ 10.13039/501100011033/ (PID2022-14044NB-C21), by Junta de Andaluc{\'\i}a (FQM 101) and by 
Uni\'on Europea-NextGenerationEU ($\mathrm{AST22\_8.4}$).

\end{document}